\documentstyle[12pt]{article}
\newcommand{\dbs}{\renewcommand{\baselinestretch}{1.5}
\large\normalsize}
\dbs
\begin{document}
\title{
Generalized Stacking Fault Energy Surfaces and
Dislocation Properties of Silicon: A First-Principles Theoretical
Study
}
\author{
Yu-Min Juan
and Efthimios Kaxiras\\
%}
%\address{
Department of Physics and Division of Applied Sciences \\
Harvard University, Cambridge MA 02138\\
}
\maketitle
\begin{abstract}
The generalized stacking fault (GSF) energy surfaces 
have received considerable attention due to their close relation 
to the mechanical properties of solids.
We present a detailed study of the 
GSF energy surfaces of silicon within the framework
of density functional theory.
We have calculated the GSF energy surfaces for
the shuffle and glide set of the (111) plane, and that of the (100)
plane of silicon, paying particular attention
to the effects of the relaxation of atomic coordinates. 
Based on the calculated GSF energy surfaces and
the Peierls-Nabarro model, we obtain
estimates for the dislocation profiles, core energies, 
Peierls energies, and the corresponding stresses
for various planar dislocations of silicon.
\end{abstract}

%\dbs

\section{Introduction}
\label{pnintroduction}
An issue of central importance in materials
science is the 
intrinsic ductility or brittleness of solids. 
To address this question, one has to consider how a crack
in the solid will
respond to external loading.
Since dislocation nucleation at a crack tip will cause the tip to blunt, 
it is customary  to associate this process with ductile
behavior.
In contrast,
brittle behavior is associated with crack propagation without dislocation
emission,
corresponding to the creation of a sharp crack tip.
Therefore, the intrinsic ductility or brittleness of a solid can be determined 
by comparing
the likelihood for dislocation nucleation at the crack tip
to Griffith's criterion for cleavage (Griffith (1920)).  

Estimating the likelihood of dislocation
nucleation at a crack tip under external
loading is a non-trivial task. Significant advances have been made 
by Rice and collaborators recently 
in modeling dislocation
nucleation at a crack tip based on the Peierls stress concept
(Rice (1992), 
Beltz and Rice (1991, 1992), Rice and Beltz (1994),
Rice, Beltz, and Sun (1992), Beltz (1992), Sun, Beltz, and Rice (1993))
In that work,  
an important solid state parameter, the unstable stacking fault
energy denoted by $\gamma_{us}$, was identified 
as the controlling parameter for dislocation emission
at a crack tip under shear loading. 
One of the attractive features of Rice's theory is that 
the value of $\gamma_{us}$ can be obtained from theoretical
calculations without any ambiguity:
Consider the process by which an infinite crystal is cut in half
along a plane, and the upper half is sheared with respect to the lower 
half by a displacement vector $\vec{f}$ and let
$\Phi(\vec{f})$ be  the energy per unit area on
the slip plane associated with this displacement.
The energy surface $\Phi(\vec{f})$ 
obtained as a
function of the generalized displacement vector $\vec{f}$ is called
the generalized
stacking fault (GSF) energy surface.
The GSF energy surface or $\gamma$ surface and its 
relation to mechanical properties of solids were first considered by 
Vitek (1966), (1967), (1968), 
Vitek and Yamaguchi (1973), and Yamaguchi and Vitek (1975). 
In this context,   
Rice's unstable stacking energy 
$\gamma_{us}$ is the lowest energy barrier in the $\gamma$ surface
that has to be surmounted
during the shearing process that takes a crystal 
from an ideal configuration to another equivalent one.

In addition to its relevance to dislocation emission, 
the GSF energy surface can be used to obtain
the restoring force due to the misfit of the lattice near the 
core of a dislocation (see for example, J\`{o}os, Ren, and Duesbery (1994)). 
By incorporating these restoring forces 
into the Peierls-Nabarro model (Peierls (1940), Nabarro (1947)), 
one can calculate dislocation profiles, core energies, Peierls energies,
and Peierls stresses for planar dislocations on the corresponding
crystal plane. 
The calculation of dislocation properties within the Peierls-Nabarro
model represents a drastic simplification, because 
it neglects the discreteness of the lattice by treating
the problem in a continuum picture. 
Despite its obvious limitations, 
this approach provides a useful phenomenological framework for 
comparison of the
structural properties as well as the energetics of various dislocations.

From the above discussion, it is evident that
an accurate description
of the GSF energy surface is desirable. 
Empirical methods based on classical interatomic potentials
are not sufficiently accurate for this task 
(Duesbery, Michel, Kaxiras, and J\`{o}os (1991)). 
Given the significance of GSF energy surfaces, it is important
to obtain these values from
a first-principles theoretical
point of view, i.e. 
from calculations which are free of adjustable parameters. 
First-principles calculations based on 
Density Functional Theory (DFT) 
(Hohenberg and Kohn (1964), Kohn and Sham (1965)), which are
computationally much more demanding than empirical approaches, 
have been shown to be very successful in predicting
the energy differences between different structures of solids.
We have performed first-principles DFT
calculations for both the shuffle and glide sets of
the (111) plane as well as the (100) plane of silicon in the diamond lattice.
There are several reasons for choosing these particular
planes.  These are low index planes of this crystal system, 
with in-plane dense packing of the atoms and relatively large 
spacing between planes.  
Therefore, these planes are natural candidates to be exposed during cleavage.
Moreover, the Burgers vectors associated with dislocations on these planes
are short compared to other planes, which suggests a smaller
dislocation core energy.

The remaining of this paper is organized as follows:
Section~\ref{pnmeth} describes the computational techniques used in
our first-principles calculations for GSF energy surfaces.
Section~\ref{energetics} contains
our results for the
GSF energy surfaces of the (111) and (100) planes of silicon. The effects due to
the relaxation of atomic coordinates are also discussed there.
In Section~\ref{pnsec} we give a brief review of the Peierls-Nabarro
model and then present the dislocation properties obtained within
this model from the calculated GSF energy surfaces. 
We conclude with some remarks on the applicability of  
Peierls-Nabarro model in Section~\ref{summary}.

\section{Computational Methods}
\label{pnmeth}
The local density approximation (LDA) 
to the exchange-correlation functional of DFT 
proposed by Perdew and Zunger (1984) was used for the GSF 
energy calculations.
The valence electron wave functions were expanded in 
a plane wave basis. The highest kinetic energy of the plane
waves included in the basis set is 8 Ry.
The ionic
potential, including the screening by core electrons, 
was modeled by a nonlocal
norm-conserving pseudopotential 
from Bachelet, Greenside, Barraff, and Schl\"{u}ter (1982),  
and the scheme  
of Kleinman and Bylander (1982) 
was employed to make the
potential separable in Fourier space.
The $d$ angular-momentum component 
was treated as the local part of the
potential with the $s$ and $p$ components containing 
the nonlocal contributions.
To simulate the block shearing process,
a supercell consisting of 12 atomic planes in the direction perpendicular
to the cut was
used in these calculations.
For the reciprocal space integration, we have used 20 special k-points in the 
irreducible Brillouin zone,
using the scheme of Monkhorst and Pack (1976). 
The approach of Car and Parrinello (1985),
which allows simultaneous
relaxation of both the ionic and the electronic degrees of freedom, was
employed in the present calculations. 
The minimum energy was obtained with the steepest descent method.
Atoms on the four atomic planes farthest from
the plane of the cut were
kept frozen to simulate the
effects of the bulk. 
Atomic relaxations are taken into account my minimizing the 
magnitude of forces calculated from the Hellmann-Feynman theorem.
The structures are considered
fully relaxed when the magnitude of the forces is smaller
than 0.005 Ry/a.u.

The unrelaxed GSF energy surface is obtained by 
simply moving one half of the crystal
rigidly with respect to the other half.
This GSF energy surface  
can be altered when atoms on either side of the cut
are allowed to respond to shearing forces. 
The difference in GSF energy surfaces with and without atomic 
relaxation can be very important, and can even change siginificantly the 
magnitude of $\gamma_{us}$ 
as well as the configuration corresponding to it (see next section). 

There is one subtlety that needs 
to be clarified in relation to the relaxation of ionic coordinates.
Within Rice's continuum theory (Rice (1992))
the GSF energy is defined as a function of the displacement 
on either side  
of a mathematical cut in the middle of the crystal.
In a real crystal, the closest approximation to this displacement  
is the relative displacement $\vec{f}$ of the two atomic planes 
immediately adjacent to the cut. 
When the two blocks on either side of the cut are moved rigidly,
$\vec{f}$ is identical to the relative displacement  
of the centers of the two halves of the crystal.
When relaxation is included,
special care must be taken in interpreting the proper
value of $\vec{f}$ from the in-plane coordinates of the ions.
For certain values of $\vec{f}$,
where $\Phi(\vec{f})$ is required by symmetry to have an extremum
(as in the case of $\gamma_{us}$), only
relaxations that are perpendicular to the plane of the cut are allowed.
In these cases, 
$\vec{f}$ is again identified with the relative displacement  
of the centers of the two halves of the crystal.
When symmetry constraints are absent, atomic relaxations in all
directions are allowed.
In these cases, the correct value of the relative 
displacement $\vec{f}$ 
is obtained from the relaxed in-plane coordinates 
of atoms on the two planes 
immediately adjacent to the cut, rather than  from the relative displacement
of the centers of the two halves of the crystal.
Therefore, for a given relative displacement $\vec{f}_C$ 
between the centers of the 
two halves of the crystal, atomic relaxation not only reduces the 
energy but also defines the actual displacement $\vec{f}$ 
(generally $\vec{f} \neq \vec{f}_C$),
which is the value relevant to the GSF energy surface.

\section{GSF Energies}
\label{energetics}
We first consider the (111) plane which is the natural cleavage
plane of silicon.
For this plane of the diamond lattice, 
there are two distinct ways to cut
the crystal: the shuffle set and the glide set. 
The difference between these two ways
is shown in Fig.\ 1(a).
For the shuffle set, the vertical distance between the two adjacent
atomic planes immediately above and below the cut is equal to the bond length;
only one bond per atom on either
side of the cut is broken during the block shearing
process. 
The corresponding interplanar
distance for the glide set
is only 1/3 of the bond length; three bonds per atom on
either side of the cut are broken
in the slip along this plane.
The calculated GSF energy surfaces for the corresponding cuts
in rigid block shearing 
with all the atoms in the two respective half 
crystals held fixed during 
shearing, are shown in Fig.\ 2(a) and 3(a) respectively.
The effects of atomic relaxation on the GSF energy
surfaces are shown in Fig.\ 2(b) and 3(b)
for the shuffle and glide sets. 
Because the distance between the two atomic planes
adjacent to the cut is sufficiently large in the shuffle set,
the relaxation effects on this energy surface are very small and
hardly noticeable on the scale of $\gamma_{us}$, which is of order 2 eV
[compare Fig.\ 2(a) and 2(b)].

The effect of atomic relaxation is much more pronounced in the glide
set.
In Fig.\ 1 we show the atomic positions on a (110) plane, during the shearing
on the glide plane along the $<1\bar{2}1>$ direction, for {\em unrelaxed}
structures.
Fig.\ \ref{strpath}(a) is the original ideal structure. 
Fig.\ \ref{strpath}(b) is the structure for the {\em unstable} stacking fault
configuration.
Fig.\ \ref{strpath}(c) is the {\em stable}
stacking fault configuration
for the glide set. 
There is essentially
no difference 
between the stable stacking fault structure and
the ideal structure,
as far as the coordination numbers of the two sets 
of atoms adjacent to the cut are concerned.
For this reason, the energy of the stable stacking fault 
configuration is very low (see Fig. 3).
For displacements beyond the stable stacking fault 
configuration along the $<1\bar{2}1>$ direction, the atoms belonging
to the same color (white or black)
immediately adjacent to the cut start getting very close to each
other as can be seen from 
Fig.\ \ref{strpath}(d), (e)  and (f). 
Relaxation is restricted by symmetry to the direction 
perpendicular to the plane of the cut for configurations 
~\ref{strpath}(b), (c) and (e).
Especially in Fig.\ \ref{strpath}(e), the
atoms are within a very small distance, almost on top of each other.
This kind of arrangement of atoms costs a significant amount
of energy due to the strong repulsive interaction between the ionic cores.
Configuration~\ref{strpath}(e) 
actually corresponds to the highest energy for the glide cut.
The {\em relaxed} energy surface
on the glide set has the same energy scale as that for the shuffle set. 
The reason for
this dramatic reduction in energy [compare Fig.\ 3(a) and 3(b)]
is that through relaxation
the atoms can avoid coming very close
to each other during the sliding process.

The results for the (100) plane before and after atomic 
relaxation of the atomic coordinates are shown in 
Fig.\ 4(a) and 4(b) respectively. The origin for the high
energy barrier along the $<$01$\bar{1}$$>$ path is similar to the situation for
the glide set, where atoms adjacent to the cut come very close to each other.
Therefore, the relaxation of the atomic coordinates is expected to
reduce the energy barrier along this direction significantly. 
As is shown in Fig.\ 4, the topology of the
energy surface is different after atoms 
are allowed to relax. One point deserves further attention:
the $<$01$\bar{1}$$>$ path, which is the path containing the highest
energy barrier before relaxation, has become the 
energetically favorable path for the 
gliding process on the (100) plane after taking into account the effects 
of atomic relaxation.

The calculated values of
$\gamma_{us}$ both before and after the 
relaxation of atomic coordinates for the (111) and (100)
planes of silicon are summarized in Table I. 
The expressions that give
the relaxed
GSF energy surfaces, fitted by sinusoidal expansions,
are given in the Appendix.

Based on Rice's theory, a direct comparison
of the values of $\gamma_{us}$
indicates that it is energetically more
favorable to nucleate dislocations on the shuffle set of the (111) plane
under zero temperature and zero pressure conditions.
The changes in the energy surface due to atomic relaxation will also
affect the estimates of the free energy.
Following Kaxiras and Duesbery (1993), we define the free
energy per unit area associated with a particular slip process as:
\begin{equation}
F=\gamma_{us}-T \frac {S}{A}-P \frac {\Delta V}{A},
\end{equation}
where $S$ is the entropy and $\Delta V$ the volume relaxation at
the saddle point.
$\Delta V$ is obtained by minimizing the energy of the 
saddle point configuration with respect to the magnitude 
of the slab thickness. 
We use Vineyard's transition state theory (Vineyard (1957))  as
outlined by Kaxiras and Duesbery (1993), to estimate the
entropy $S$ for the relaxed energy surface. The condition that 
the preferred slip plane changes from shuffle to glide
is then given by $F^{glide}=F^{shuffle}$. The $(P,T)$ values that satisfy 
this condition are shown in 
Fig.\ \ref{entropy}. 
As anticipated in the work of Kaxiras and Duesbery (1993), 
the larger reduction in the energy 
associated with relaxation of 
the glide cut results in  an overall shift of the
$F^{glide}=F^{shuffle}$ line towards higher 
pressure and lower temperature, compared to the unrelaxed results. 
For example, the stress needed to change from the shuffle to the
glide set at room
temperature (300$^{0}$ K)
corresponds to 
an external pressure of only 15 kbar as calculated
from the results with the relaxed energy surface, 
while a pressure as high as 53 kbar would be predicted
if the unrelaxed energy surface were used. 

However, even the results obtained from the relaxed energy surface
calculations should not be taken literally for the following reasons:
(1) Other
factors (such as the electronic
degrees of freedom), which  are neglected in the present
entropy calculation, 
could make a significant contribution to the free energy. 
(2) The comparison of free energies is based
on Rice's theory in which 
the energy associated with the creation of surface
during dislocation 
emission is neglected; further 
investigations
are needed to get an estimate of such effects (see for instance
Xu, Argon and Ortiz (1995)).
A detailed discussion on surface effects in dislocation 
nucleation for silicon will be published 
elsewhere (Juan, Kaxiras, and Sun (1995)).
(3) Effects related to reconstruction of the dislocation core
are also neglected in this theoretical framework.  For a stiff covalent 
material, core reconstruction effects are very important 
(see Bulatov, Yip and Argon (1995)), and are expected to 
change the values of $\gamma_{us}$.
All these approximations may affect the exact numbers for the transition
from shuffle to glide dominance, 
even though we expect the qualitative picture to remain the same.

\section{Peierls-Nabarro Model for Planar Dislocation}
\label{pnsec}
By combining our results for the GSF energy surfaces
with the Peierls-Nabarro model (Peierls (1940), Nabarro (1947)), we can further analyze
the properties of dislocations on these planes within continuum elastic
theory. We give here a brief description of 
the Peierls-Nabarro model 
which is based on the premise that a
dislocation can be thought of as a continuous
distribution of infinitesimal crystal misfits on the glide plane 
(Kroupa and Lej\v{c}ek (1972), Hirth and Lothe (1982), 
J\`{o}os et al.\ (1994)). 
In the following we treat the distortion due to the dislocation 
as a scalar function of position $f(x)$ for simplicity, although the 
true vectorial character of $\vec{f}$ should be considered 
in a general treatment. 
For each point on the glide plane at
a distance $x'$ from the dislocation line, there is a corresponding 
infinitesimal Burgers vector $df'= df/dx|_{x'} dx'$,
where $df/dx$ is defined to be the dislocation density $\rho(x)$
and $f(x)$ is the disregistry at the point $x$, satisfying the condition:
\begin{equation}
\int_{-\infty}^{+\infty}\rho(x')dx'=
\int_{-\infty}^{+\infty}\frac{df(x')}{dx'}dx'
=b,
\end{equation}
with $b$ the Burgers vector of the dislocation. 
From the elastic model, the stress due to a dislocation 
along the direction of the Burgers vector on the slip plane is given by 
\begin{equation}
\sigma_{bn}=\frac{K}{2\pi} \frac{b}{x},
\end{equation}
where $x$ is the distance between the dislocation and the point
where the stress is evaluated,
and $n$ is the normal
to the glide plane. The constant $K$ 
depends on the elastic properties of the crystal and the 
dislocation type. Its value for an isotropic solid 
is given by J\`{o}os et al.\  (1994):
\begin{equation}
K=\mu\left[\frac{\sin^{2}(\theta)}{(1-\nu)}+\cos^{2}(\theta)\right], 
\label{Kdef}
\end{equation}
where $\mu$ and $\nu$ are the shear modulus and Poisson's ratio
respectively, and $\theta$ is the angle between the dislocation 
line and the Burgers vector.
Therefore, the stress due to the existence of the dislocation
for a point $x$ on the glide plane is:
\begin{equation}
\frac{K}{2\pi}\int_{-\infty}^{+\infty}\frac{\rho(x')}{x-x'} dx'.
\end{equation}
Meanwhile, there is another stress which is due to the
local disregistry of the crystal and tends to restore
the crystal. This crystal restoring stress is a periodic function of the
displacement and can be represented as $F_{b}(f)$.
Equilibrium is attained when the stresses due to the two
terms are in balance with each other
as expressed in the following equation, known as
the Peierls-Nabarro equation,
\begin{equation}
\frac{K}{2\pi}\int_{-\infty}^{+\infty}\frac{\rho(x')}{x-x'} dx'=F_{b}(f(x)).
\end{equation}
The dislocation profile $f(x)$ and the dislocation density 
$\rho(x) = df/dx$ can be determined
by solving this equation
with the normalization condition $\int_{-\infty}^{+\infty}\rho(x)dx=b$. 
In the original Peierls-Nabarro model, a simple sinusoidal form is 
assumed for the restoring stress,
\begin{equation}
F_{b}(f(x))=F_{max} \sin (\frac {2 \pi f(x)} {b}), 
\end{equation}
with $F_{max}$ the maximum
stress. This assumption leads to the analytic solution 
\begin{equation}
f(x)=\frac {b} {\pi} \tan ^{-1} \frac {x}{\zeta} +\frac {b}{2}.
\end{equation}
The parameter  $\zeta=Kb/4 \pi F_{max}$ can be viewed as 
the width of the dislocation within this model,
since the value of the dislocation density $\rho$ at this point is 
exactly one half its value at $x=0$.
Within the GSF approach
the crystal restoring force $F_{b}(f)$ can be obtained by 
simply taking the derivative of
the GSF energy surface $\gamma (f)$ 
with respect to the lattice distortion $f$ (Kroupa and Lej\v{c}ek (1972),
Vitek and Yamaguchi (1973), Yamaguchi and Vitek (1975)).
For a general functional form of $F$, there is no analytic solution
for $f(x)$. 
We follow the procedure of Jo\'{o}s  et al.\ (1994)
to solve the Peierls-Nabarro
equation numerically.
This is done by expressing the disregistry vector $f(x)$ 
as a series
\begin{equation}
f(x)=\frac{b} {\pi} \sum _{i=1}^{n} \alpha_{i} \tan ^{-1} 
\frac {x-x_{i}} {\zeta_{i}} +\frac{b}{2}, 
\label{eq:expan}
\end{equation}
with the parameters $\alpha_{i}$, $\zeta_{i}$, $x_{i}$ to be determined,
subject to the normalization condition 
\begin{equation}
\sum_{i=1}^{n}\alpha_{i}=1,
\end{equation}
since
the total Burgers vector should be equal to $b$.
By substituting the above displacement formula into the two
sides of the Peierls-Nabarro equation, we get the forces 
$F(\alpha_{i}, \zeta_{i}, x_{i}, x)$ and 
$F'(\alpha_{i}, \zeta_{i}, x_{i}, x)$ 
corresponding
to the contributions from dislocation distribution and crystal 
distortion respectively.
The difference between these two forces 
$|F(\alpha_{i}, \zeta_{i}, x_{i}, x)- F'(\alpha_{i}, \zeta_{i}, x_{i}, x)|$ 
is then minimized by varying the parameters $\alpha_{i}$, $\zeta_{i}$ 
and $x_{i}$ so that the Peierls-Nabarro equation is satisfied numerically.
We find that a numerical solution is feasible by retaining only three
terms in Eq. (\ref{eq:expan}).

We display in Fig.\ \ref{enefor} the GSF energies and forces 
obtained from our first-principles
calculations with the full relaxation of the atomic 
coordinates, along the directions which are relevant to the dislocations
we will consider.
For the case of the $<1\bar{2}1>$ direction on the glide plane, only
the portion ranging from the ideal structure to the stacking fault
configuration is shown, since a
partial dislocation will be formed along this direction.
One feature of these curves deserves further comment:
not all the restoring forces can be well described
by a simple sinusoidal from (see also the recent work of Xu, Argon and
Ortiz (1995)). 
Especially in the case of the (100) plane, the force curve is very
flat when the displacement is approximately one half of
the full Burgers vector.
The significance of this deviation will be discussed in detail later, when
we consider the profiles of various dislocations. 

The disregistry vector for full dislocations
on the shuffle set of the (111) plane, 
the glide set of the (111) plane,
and the (100) plane
are displayed in Fig.\ \ref{disreg} (a), (b), and (c)
respectively. We note that
the dislocations with Burgers vector along the $<10\bar{1}>$ direction
on the glide plane (glide-60$^{0}$ and glide-screw) are 
more concentrated  compared to the
other dislocations on the (111) plane
(this effect will become more obvious when we compare the dislocation density). 
This difference in the distribution density can be understood
by considering
the magnitude of the crystal restoring force 
for these dislocations.
A qualitative estimate can be obtained by 
considering the half
width of the dislocation $\zeta$ within the classical model,
which is inversely proportional to the magnitude of the crystal
restoring force.
This force is largest in the $<10\bar{1}>$ direction of the 
glide plane, as is evident by comparing Figures 2, 3 and 4.

In Fig.\ \ref{disdensity} we show
representative dislocation densities for the planes we have considered,
including (a) the 60$^{0}$ dislocation 
on the shuffle set, (b) the 60$^{0}$ dislocation on the glide set,
and (c) the 90$^{0}$ dislocation
on the (100) plane.
We have displayed
the results obtained from both the relaxed (upper panel) and the unrelaxed
GSF energy surfaces (lower panel) to examine the effects due to the relaxation
of the atomic coordinates. For comparison we have also displayed
the results from the classical solution (dashed curves),
obtained by assuming a
sinusoidal form for the crystal restoring force with its maximum
equal to the value obtained from the first-principles calculations.
For the case of the 60$^{0}$ dislocation on the shuffle set, 
it is apparent that the use of the relaxed
GSF energy surface does not cause significant change on the dislocation
profile both qualitatively and quantitatively. This
is a consequence of the fact that atomic relaxation 
does not change the GSF energies along this direction
in any significant way.
For the 60$^{0}$ dislocation on the glide set, the use of the relaxed
energy surface makes the dislocation profile considerably smoother and wider
but the general shape remains unchanged.
This is due to the fact
that the relaxation reduces the
magnitude of the crystal restoring force  but does not
change its functional form significantly.
For the case
of the 90$^{0}$ dislocation on the (100) plane, the dislocation
profile is changed {\em qualitatively} with the use of the relaxed
energy surface. This reflects the fact that the crystal restoring
force for this plane is significantly different from the sinusoidal 
function as we mentioned above.
The double peak indicates the dissociation of the
full dislocation into two fractional
dislocations (Vitek and Kroupa (1969)). 
A brief summary of the obtained dislocation properties 
is given in Table II.

Another useful quantity to consider is the 
energy barrier 
associated with dislocation motion.  Within the
Peierls-Nabarro model, this energy barrier is called the Peierls energy
and is defined as the amplitude of the variation
of the misfit energy on the glide plane as the position
of the dislocation line moves.
With the obtained dislocation profiles, we can calculate
the misfit energy 
across the glide plane as a function of the
position of the dislocation line $u$ following 
the definition given by 
Vitek and Yamaguchi (1973) and J\'{o}os et al.\ (1994).
To be consistent with the Peierls-Nabarro model, where
the discreteness
of the crystal is taken into account at the glide plane,
the misfit energy $W(u)$ is defined as the sum of
all the 
misfit energies between pairs of atom rows, obtained 
from the GSF energy at the
local disregistry:
\begin{equation}
W(u)= \sum_{m=-\infty}^{+\infty} \gamma (f(ma'-u)) a',
\end{equation}
where $a'$ is the distance between adjacent atomic planes
in the direction perpendicular to the dislocation line. 
This expression meets
two important requirements:
First, it has the correct period $a'$ of planes in the crystal,
\begin{equation}
W(u+a')=W(u). 
\end{equation}
Second, in  the limit
of a very narrow dislocation, it reproduces the correct
maximum $\gamma_{us} a'$.
The other quantity which might be of interest is $W(a'/2)$ where the
minimum of the misfit energy function occurs. 
Since this quantity 
measures the nonelastic part of the energy of the dislocation, it
provides a qualitative estimate of the core energy.
The stress associated with the energy function $W(u)$ can then be defined as:
\begin{equation}
\sigma(u)=\frac{1}{a'}\frac{dW(u')}{du'}|_{u'=u}.
\end{equation}
The maximum of this stress function is the Peierls stress,
needed to move the dislocation.
The values of the energy function $W(u)$ and the stress function $\sigma(u)$
for the (111) plane are shown 
in Fig.\ \ref{wu1} (a) and (b). It is obvious
that the dislocation with Burgers vector along the $<10\bar{1}>$ direction
on the glide set (labeled Glide-60$^{0}$ in Fig.\ \ref{wu1} (b))
has larger Peierls energy and Peierls stress.
This is expected since we know
from our GSF energy surface calculations that the energy barrier along
that direction is higher than those along the other two vectors. 
Fig.\ \ref{wu1}(c) is
the energy and stress function for the case of the (100) plane.
The different behavior between the 90$^{0}$ and the screw dislocations
on the (100) plane 
is primarily due to the difference in the values of the constant $K$. 
The Peierls energy, stress, and core energy of
various dislocations are summarized in Table III.
For the (111) plane, it appears that the dislocations 
belonging to the shuffle set
are the easiest to move under zero temperature and zero pressure
conditions. However, as we discussed earlier, including 
the effects of temperature, pressure, surface creation and 
core reconstruction can change the picture.
Finally, we would like
to mention that there is experimental evidence that on the glide set
the 90$^{0}$ partial is more mobile than the 30$^{0}$ partial 
(Wessel and Alexander (1977), 
Alexander, Gottschalk, and Kisielowski-Kemmerich (1985), 
Grosbras, Demenet, Garem, and Desoyer (1984), 
Demenet, Grosbras, Garem, and Desoyer (1989)),
which is consistent with the results of our calculations,
namely $\sigma(u)$ is lower for the glide-90$^0$(p) than the
glide-30$^0$(p) dislocation, see Fig. 9(b).

\section{Summary}
\label{summary}
In summary, we have performed first-principles calculations to obtain
the GSF energy surfaces for both the shuffle and glide sets of the (111) plane,
as well as the (100) plane of silicon. We showed that for the glide set
and the (100) plane, the effects on the
GSF energy surfaces due to the relaxation of the atomic
coordinates are significant. The unstable stacking fault energies
$\gamma_{us}$ for these planes were determined from our calculations.
By combining these values  with Rice's criterion for dislocation
nucleation, the shuffle set appears to be favored for
dislocation emission under
zero temperature and zero pressure conditions. 
A qualitative account of entropy effects was attempted, 
based on Vineyard's transition state theory.
By comparing the free energies of the shuffle and glide sets,
we find that either set can
dominate under different thermodynamical conditions. 
Quantitative free-energy comparisons should also take into account 
additional entropy effects, and the effects of 
surface creation and core reconstruction.

The GSF energy surfaces were then combined with the Peierls-Nabarro
model to investigate the dislocation properties of silicon. The
crystal restoring forces are obtained directly from our first-principles
GSF energy surfaces. We demonstrated the importance
of using the relaxed energy surface for the calculation of these properties.
The differences between dislocation profiles 
obtained with relaxed GSF and unrelaxed GSF
energy surface are significant in certain cases. 
The Peierls energy and Peierls stress were also calculated within
the framework of the Peierls-Nabarro model. 

We would like to conclude with a brief discussion on the approximations involved
in the Peierls-Nabarro model: (1) The
dislocation is assumed to be planar within this model.
We expect that the errors introduced with this approximation
for a stiff material like silicon should be minimal. (2) The response from
the crystal is treated within elastic theory, which may not apply
for a very narrow core situation, where non-elastic effects
are expected to be important. (3) The Peierls stress
concept assumes that the only relevant quantity in determining
the stress is the local disregistry, which is true only
when the dislocation density is very smooth. (4) The dislocation line 
is assumed 
to move as a rigid object, which is unlikely (see e.g. the recent 
work of Bulatov, Yip and Argon (1995) on this issue).  Accordingly, 
we suggest that these
results should not be taken literally.  Rather, our  
investigation provides a {\em qualitative}
comparative discussion of dislocation properties in silicon, that can 
serve as guide to more detailed studies.

\section{Acknowledgement}
We are indebted to J.R. Rice and his coworkers Y. Sun and G. Beltz for
useful discussions and insight on GSF theory. 
We acknowledge useful discussions with M.S. Duesbery,
B. J\`{o}os, A. Argon and S. Yip on dislocation properties.
We are particularly thankful to V. Bulatov for a critical 
reading of the manuscript and many useful suggestions.

This work was
supported in part by ONR grant \#N00014-92-J-1960 and in part by Harvard's 
Materials Research Science and Engineering Center, which is funded through NSF.
The calculations were performed at the 
Department of Defense NAVOCEANO and CEWES
computational facilities.
\newpage

\section*{Appendix}
We used sinusoidal expansions to fit the calculated GSF energy surfaces 
in order to facilitate the computation of dislocation properties. 
The expansions were chosen so that they satisfy the underlying 
translational symmetries of the lattice.
We have checked the numerical values and found that 
the numbers obtained from the fittng formula
reflect the underlying rotational symmetry with sufficient accuracy.
Specifically, we use the following expression:
\begin{eqnarray}
\gamma(x,y)&=&\sum_{n,m}  A_{nm}\cos (\frac {2 \pi n}{a_1} x)
\cos (\frac {2 \pi m}{a_2}y) 
+ B_{nm}\cos 
(\frac {2 \pi n}{a_1}x) \sin (\frac {2 \pi m}{a_2}y) \nonumber \\
& &+C_{nm}\sin(\frac {2 \pi n}{a_1}x) \sin(\frac {2 \pi m}{a_2}y)
\end{eqnarray}
where $a_1$ and $a_2$ are the repeat distances in the $(x,y)$ plane.
In terms of the lattice constant $a_{0}$ of bulk Si, these are given by 
$a_1=a_{0} / \sqrt{2}$, $a_2=a_{0}\times \sqrt{3/2}$
for the (111) plane cuts (both shuffle and glide), and 
$a_1 = a_2 = a_{0}/ \sqrt{2}$ for the (100) plane.
The $(x,y)$ directions in the (111) plane cuts correspond 
to the $<10\bar{1}>$ and 
$<1\bar{2}1>$ crystallographic directions, whereas in the (100) plane cut they
correspond to the $<011>$ and $<01\bar{1}>$ crystallographic directions.
The coefficients of the  terms retained in the above expansion are given in
Table IV.

\vspace{0.6in}

\section*{References}

Alexander H., Gottschalk H., and  Kisielowski-Kemmerich C., (1985)
in {\it Dislocations
in Solids}, edited by H. Suzuki, T.Minomoya, K. Sumino, and S. Takeuchi 
(University of Tokyo Press, Tokyo).

Bachelet G., Greenside H.,  Barraff G., and Schl\"{u}ter M., (1981) Phys. Rev.
{\bf 24}, 4745.

Beltz G.E., (1992) Ph.D Thesis, Division of Applied Sciences,
Harvard University, Cambridge, MA.

Beltz G.E. and Rice J.R., (1991) in {\it Modeling the Deformation
of Crystalline Solids: Physical Theory, Applications, and
Experimental Comparisons}, edited by T.C. Lowe, A.D. Rollett, P.S. 
Follansbee and G.S. Dehn, (TMS Minerals, Metals and Materials Society,
Warrendale).

Beltz G.E. and Rice J.R., (1992) Acta Metall. {\bf 40} S321.

Bulatov V.V., Yip S., and Argon, A.S. (1995) Phil. Mag. A {\bf 72}, 453. 

Car R. and Parrinello M., (1985) Phys. Rev. Lett. {\bf 55}, 2471.

Demenet J.L., Grosbras P., Garem H., and  Desoyer J.C., (1989)
Phil. Mag. A {\bf 59}, 501.

Duesbery M.S., Michel D.J., Kaxiras E., and J\`{o}os B., (1991) Mat. Res. Soc. Symp. Proc.
Vol. 209, p. 125.

Griffith A.A., (1920) Phil. Trans. R. Soc. {\bf A 184}, 181. 

Grosbras P.,  Demenet J.L., Garem H., and  Desoyer J.C., (1984)
Phys. Status Solidi A {\bf 84}, 481.

Hirth J.P.\  and Lothe J., (1982) Theory of Dislocations, 2nd ed. (Wiley, New York).

Hohenberg P.\ and Kohn W., (1964) Phys. Rev. {\bf 136}, B864.

J\`{o}os B., Ren Q., and Duesbery M.S., (1994)
Phys. Rev. B {\bf 50}, 5890.

Juan Y.M, Kaxiras E. and Sun Y., (1995) Phil. Mag. Lett. (to be published).

Kaxiras E.\ and  Duesbery M.S., (1993) Phys. Rev. Lett. {\bf 70}, 3752.

Kleinman L.\ and Bylander D.M., (1982) Phys. Rev. Lett. {\bf 48}, 1425.

Kohn W.\ and Sham L., (1965) Phys. Rev. {\bf 140}, A1133.

Kroupa F.\ and  Lej\v{c}ek L., (1972) Czech. J. Phys. Rev. B. {\bf 22}, 813.

Louchet F.\ and  Thibault-Desseaux J., (1987) Rev. Phys. Appl. {\bf 22}, 207. 

Monkhorst H.J.\ and  Pack J.D., (1976) Phys. Rev. B. {\bf 13}, 5188.

Nabarro F.R.N., (1947) Proc. Phys. Soc. London {\bf 59}, 256.

Peierls R., (1940) Proc. Phys. Soc. London {\bf 52}, 34.

Perdew J.\ and  Zunger A., (1984) Phys. Rev. B {\bf 23}, 5048.

Rice, J.R., (1992) J. Mech. Phys. Solids, {\bf 40}, 239.

Rice J.R.\ and Beltz G.E., (1994)  J. Mech. Phys. Solids, {\bf 42}, 333.

Rice J.R., Beltz G.E., and Sun Y., (1992) 
in {\it Topics in Fracture and Fatigue},
edited by A.S. Argon, (Springer, Berlin).

Sun Y., Beltz G.E., and Rice J.R., (1993)
Mater. Sci. Engng {\bf A170} 67.

Vineyard G.H., (1957) J. Phys. Chem. Solids {\bf 3}, 121.

Vitek V., (1966) Phys. Stat. Sol. {\bf 18}, 683. 

Vitek V., (1967) Phys. Stat. Sol. {\bf 22}, 453. 

Vitek V., (1968) Phil. Mag. {\bf 18}, 773.

Vitek V.\ and and Kroupa F., Phil. Mag. {\bf 19}, 265.

Vitek V.\ and Yamaguchi M., (1973) J. Phys. F {\bf 3}, 537.

Wessel K.\ and Alexander H., (1977) Phil. Mag.
{\bf 35}, 1523.

Yamaguchi M.\ and Vitek V., (1975) J. Phys. F {\bf 5}, 11.

Xu G., Argon A.S. and Ortiz M. (1995), Phil. Mag. A {\bf 72}, 415.

\pagebreak

\begin{center}
{\large Table I}

\vspace{0.25in}

\begin{tabular}{||c|c|c|c|c||} \hline
& \multicolumn{2}{c|}{No Relaxation}&\multicolumn{2}{c|}{Atomic Relaxation}  \\
& \multicolumn{2}{c|}{ }            &\multicolumn{2}{c|}{at ideal volume}    \\ \hline \hline
& location &$\gamma_{us}$ (J/m$^{2}$)&location&$\gamma_{us}$ (J/m$^{2}$)\\ \hline \hline
(111)-Shuffle&$\frac{1}{4} [10\bar{1}]$& 1.84&$\frac{1}{4} [10\bar{1}]$& 1.81  \\ \hline
(111)-Glide  &$\frac{1}{12}[1\bar{2}1]$&2.51 & $\frac{1}{12} [1\bar{2}1]$& 2.02 \\ \hline
(100) &$\frac{1}{4}[011]$& 2.97& $\frac{1}{4}[0\bar{1}1]$ &2.15 \\ \hline \hline
\end{tabular}
\end{center}

TABLE I: The unstable stacking fault energy, $\gamma_{us}$, 
obtained for the shuffle set
and glide set of the (111) plane, as well as the (100) plane of silicon.
The location indicates the position of relative displacement
$\vec{f}$ where the $\gamma_{us}$ occurs.

\vspace{0.35in}

\begin{center}
{\large Table II}

\vspace{0.25in}

\begin{tabular}{||c|c|c|c|c|c|c|c|c||}  \hline \hline
dislocation& \multicolumn{2}{c|}{Shuffle}&  \multicolumn{2}{c|}{Glide}& 
\multicolumn{2}{c|}{Glide partial}&  \multicolumn{2}{c||}{(100)}\\ \cline{2-9}
& 60$^{0}$&Screw&60$^{0}$&Screw&30$^{0}$&90$^{0}$&90$^{0}$&Screw\\ \hline \hline
$K (10^{11}\frac{dyne}{cm^{2}})$&8.02&6.37&8.02&6.37&6.92&8.58&9.04&6.375 \\ \hline
$\tau_{max} (10^{11} \frac {dyne} {cm^{2}})$&1.49&1.49&4.29&4.29&2.78&2.78&2.06&
2.06 \\ \hline
$\zeta$ (\AA)& 1.08&0.92&0.46&0.37&0.77&0.92&2.15&1.54 \\ \hline
b (\AA)&3.84 &3.84 &3.84 &3.84& 2.22 &2.22 &3.84 &3.84 \\ \hline \hline
\end{tabular}
\end{center}

Table II: Quantities related to the properties of dislocations
for the shuffle and glide sets on the (111) plane, 
as well as the (100) plane of silicon.  The meaning of $K$ is 
given in Eq.\ (\ref{Kdef}), $\tau_{max}$ 
is the maximum value of $d \gamma_{us} (\vec{f})/ d \vec{f}$ 
for the directions we are interested in, 
$\zeta$ is the half width of the dislocation, and $b$ is the Burgers vector.

\newpage

\begin{center}
{\large Table III}

\begin{tabular}{||c|c|c|c|c|c|c|c|c||}  \hline \hline
dislocation& \multicolumn{2}{c|}{Shuffle}&  \multicolumn{2}{c|}{Glide}&  
\multicolumn{2}{c|}{Glide partial}&  \multicolumn{2}{c||}{(100)}\\  \cline{2-9}
& 60$^{0}$&Screw&60$^{0}$&Screw&30$^{0}$&90$^{0}$&90$^{0}$&Screw\\ \hline \hline
$W_{p} (eV\AA)$&0.149&0.183&0.842&0.898&0.287&0.246&0.112&0.296\\ \hline
$\sigma_{p} (eV\AA^3)$&0.046&0.062&0.399&0.504&0.176&0.139&0.032&0.081 \\ \hline
$W(a'/2)$&0.467&0.338&0.345&0.238&0.174&0.246&0.549&0.274 \\ \hline \hline
\end{tabular}

\end{center}

Table III: The Peierls energy $W_{p}$ , Peierls stress $\sigma_{p}$, 
and the estimated core energy $W(a'/2)$ obtained
from the Peierls-Nabarro model for various dislocations on the (111) and (100)
planes of silicon.

\begin{center}
{\large Table IV}

\begin{tabular} {||l||r|r|r||r|r|r||r||} \hline \hline
   & \multicolumn{3}{c||}{(111)-shuffle}
   & \multicolumn{3}{c||}{(111)-glide}
   & \multicolumn{1}{c||}{(100)} \\ \hline \hline
(n,m)&$A_{nm}$&$B_{nm}$&$C_{nm}$
      &$A_{nm}$&$B_{nm}$&$C_{nm}$
      &$A_{nm}$\\ \hline \hline
(0,0)&0.9789& & &  2.2683& & & 1.8463\\ \hline
(0,1)&      & &  &       & & & -0.6728\\ \hline
(0,2)&-0.3857& &0.0625& -0.7059& &0.7024 &  \\ \hline
(0,4)&0.0180& &-0.0200& 0.2570& &-0.3473 &   \\ \hline
(1,0)&      & &       &     & &         &-0.8443\\ \hline
(1,1)&-0.7714&-0.1231 &  &-1.3609&-1.8219 &  & -0.2670 \\ \hline
(1,2)&       &    &   & & & & -0.1493  \\ \hline
(1,3)&0.0950&-0.0010& & -0.6291&0.1218 & & \\ \hline
(2,0)&0.0470& & & -0.0126& & &  0.1471 \\ \hline
(2,1)&      & & &  & & & 0.0087 \\ \hline
(2,2)&0.0350&0.0390&  &0.4412&0.0890& & -0.0485 \\ \hline
(2,4)&      &      &  &-0.2579&0.2698&  &  \\ \hline
(3,1)&-0.0180&-0.0070 & &  &  &  &  \\ \hline \hline
\end{tabular}
\end{center}

Table IV: The non-zero coefficients for the 
expansion of the $\gamma$ surface in terms
of sinusoidal functions for the shuffle and glide
(111) cuts, and for the (100) cut.

\newpage

\begin{figure}
\caption{
The atomic structures involved in the shearing process,
in which one half of the diamond crystal
is sheared rigidly with respect to the other half 
along the $<1\bar{2}1>$ direction
on the glide set of the (111) plane of silicon.
Both the glide (G) and shuffle (S) 
cuts are indicated in part (a) by dashed lines.
The two different shadings of the atoms represent two different (110)
atomic planes.
(a) the ideal structure.
(b) the unstable stacking fault
configuration. (c) the stacking fault configuration. (d), (e), and (f)
are intermediate geometries, when the upper half is sheared 
further along the $<1\bar{2}1>$
direction. (e) is the geometry with the highest energy. 
}\label{strpath}
\end{figure}

\begin{figure}
\caption{
The GSF energy surface: (a) before and (b) after relaxation,
for the shuffle set of the (111) plane.
}\label{gsf111r}
\end{figure}

\begin{figure}
\caption{
The GSF energy surface: (a) before and (b) after relaxation,
for the glide set of the (111) plane.
}\label{gsf111r2}
\end{figure}

\begin{figure}
\caption{
The GSF energy surface: (a) before and (b) after relaxation,
for the (100) plane.
}\label{gsf100}
\end{figure}

\begin{figure}
\caption{
The shuffle set / glide set transition phase diagram on the $(P,T)$ plane.
Results including atomic relaxation (solid line) and before
relaxation (dashed
line) are shown for comparison. For $(P,T)$ values
below (above) the line, the glide (shuffle) has lower unstable stacking 
free energy.
}\label{entropy}
\end{figure}

\begin{figure}
\caption{The fully relaxed
GSF energy curve and the corresponding stress
for the relevant directions on the shuffle, glide and (100) planes of silicon.
}\label{enefor}
\end{figure}

\begin{figure}
\caption {
(a) The disregistry vector obtained for dislocations
on the shuffle set of the (111) plane;
(b) the disregistry vector for dislocations on
the glide set of the (111) plane;
(c) the disregistry vector for dislocations on the (100) plane.
}\label{disreg}
\end{figure}

\begin{figure}
\caption{
The dislocation density obtained from our calculations for (a)
the 60$^{0}$ dislocation
on the shuffle set, (b) the 60$^{0}$ dislocation on the glide set,
and (c) the 90$^{0}$ dislocation
on the (100) plane.
Results from both 
relaxed (upper panel) and unrelaxed
(lower panel) GSF energy surfaces are displayed for comparison. 
The dashed curves are the classical solutions (see text). 
}\label{disdensity}
\end{figure}

\begin{figure}
\caption{
(a) The energy function $W(u)$ with respect to the displacement vector
$u$ (upper panel) and
the corresponding
stress $\sigma(u)$ as a function of $u$ (lower panel)
for dislocations belonging to shuffle set. 
(b) Same as in (a), for dislocations on the glide set.
(c) Same as in (a) for dislocations on the (100) plane.
}\label{wu1}
\end{figure}

\vfill 

\end{document}